\documentclass[pmlr]{jmlr}

\usepackage{longtable}

\usepackage{booktabs}
\usepackage[load-configurations=version-1]{siunitx} 

\makeatletter
\def\set@curr@file#1{\def\@curr@file{#1}} 
\makeatother

\theorembodyfont{\upshape}
\theoremheaderfont{\scshape}
\theorempostheader{:}
\theoremsep{\newline}

\jmlrvolume{}
\jmlryear{2020}
\jmlrworkshop{Machine Learning for Healthcare}

\newcolumntype{C}[1]{>{\centering\arraybackslash}p{#1}}
\newcolumntype{x}[1]{>{\centering\let\newline\\\arraybackslash\hspace{0pt}}p{#1}}
\usepackage{adjustbox}
\usepackage{float}

\title[Dynamically Extracting Outcome-Specific Problem Lists]{Dynamically Extracting Outcome-Specific Problem Lists from Clinical Notes with Guided Multi-Headed Attention}

\author{\Name{Justin Lovelace}$^1$
       \Email{justinlovelace@tamu.edu}\\ 
       \AND
       \Name{Nathan C. Hurley}$^1$
       \Email{natech@tamu.edu}\\ 
       \AND
       \Name{Adrian D. Haimovich}$^2$
       \Email{adrian.haimovich@yale.edu}\\ 
       \AND
       \Name{Bobak J. Mortazavi}$^{1}$
       \Email{bobakm@tamu.edu}\\ 
       \AND
      \addr $^1$Department of Computer Science and Engineering, Texas A\&M University, USA\\
      \addr $^2$Yale School of Medicine, Yale University, USA\\}

\begin{document}

\maketitle

\begin{abstract}
  Problem lists are intended to provide clinicians with a relevant summary of patient medical issues and are embedded in many electronic health record systems. Despite their importance, problem lists are often cluttered with resolved or currently irrelevant conditions. In this work, we develop a novel end-to-end framework that first extracts diagnosis and procedure information from clinical notes and subsequently uses the extracted medical problems to predict patient outcomes. This framework is both more performant and more interpretable than existing models used within the domain, achieving an AU-ROC of 0.710 for bounceback readmission and 0.869 for in-hospital mortality occurring after ICU discharge. We identify risk factors for both readmission and mortality outcomes and demonstrate that our framework can be used to develop dynamic problem lists that present clinical problems along with their quantitative importance. We conduct a qualitative user study with medical experts and demonstrate that they view the lists produced by our framework favorably and find them to be a more effective clinical decision support tool than a strong baseline.
\end{abstract}

\section{Introduction}
Problem lists are an important component of the electronic health record (EHR) that are intended to present a clear and comprehensive overview of a patient's medical problems. These lists document illnesses, injuries, and other details that may be relevant for providing patient care and are intended to allow clinicians to quickly gain an understanding of the pertinent details necessary to make informed medical decisions and provide patients with personalized care \citep{AHIMA, h11}. Despite their potential utility, there are shortcomings with problem lists in practice. One such shortcoming is that problem lists have been shown to suffer from a great deal of clutter \cite{Holmes2012}. Irrelevant or resolved conditions accumulate over time, leading to a problem list that is overwhelming and difficult for a clinician to quickly understand. This directly impairs the ability of a problem list to serve its original purpose of providing a clear and concise overview of a patient's medical condition. 

A challenge that comes with attempting to reduce clutter is that many conditions on the list may be relevant in certain situations, but contribute to clutter in others. For example, if a patient ends up in the intensive care unit (ICU), a care unit for patients with serious medical conditions, then the attending physician likely does not care about the patient's history of joint pain. That information, however, would be important for a primary care physician to follow up on during future visits. In this case, the inclusion of chronic joint pain clutters the list for the attending physician in the ICU, but removing it from the list could decrease the quality of care that the patient receives from his/her primary care physician.

In this work, we address this problem by developing a novel end-to-end framework to extract problems from the textual narrative and then utilize the extracted problems to predict the likelihood of an outcome of interest. Although our framework is generalizeable to any clinical outcome of interest, we focus on ICU readmission and patient mortality in this work to demonstrate its utility. We extract dynamic problem lists by utilizing problem extraction as an intermediate learning objective to develop an interpretable patient representation that is then used to predict the likelihood of the target outcome. By identifying the extracted problems important for the final prediction, we can produce a problem list tailored to a specific outcome of interest.

We demonstrate that this framework is both more interpretable and more performant than the current state-of-the-art work using clinical notes for the prediction of clinical outcomes \citep{analysis_attn_clinical, khadanga2019using, nyu}. Utilizing the intermediate problem list for the final outcome prediction allows clinicians to gain a clearer understanding of the model's reasoning than prior work that only highlighted important sections of the narrative. This is because our framework directly identifies clinically meaningful problems while the prior work requires a great deal of inference and guesswork on the part of the clinician to interpret what clinical signal is being represented by the highlighted text.

For example, prior work predicting the onset of heart disease found that the word ``daughter" was predictive of that outcome. The authors stated that the word usually arose in the context of the patient being brought in by their daughter which likely signaled poor health and advanced age \citep{nyu}. While this makes sense after reviewing a large number of notes, this connection is not immediately obvious and a clinician would not have the time to conduct the necessary investigation to identify such a connection. By instead directly extracting predefined clinical conditions and procedures and using those for the final prediction, we reduce the need for such inference on the part of the physician.

The primary contributions of this work are:
\vspace{-\topsep}
\begin{itemize}
  \setlength{\parskip}{0pt}
  \setlength{\itemsep}{0pt}
  \item A novel end-to-end framework for the extraction of clinical problems and the prediction of clinical outcomes that is both more interpretable and performant than models used in prior work.
  \item An expert evaluation that demonstrates that our problem extraction model exhibits robustness to labeling errors contained in a real world clinical dataset.
  \item Dynamic problem lists that report the quantitative importance of each extracted problem to an outcome of interest, providing clinicians with a concise overview of a patient's medical state and a clear understanding of the factors responsible for the model's prediction.
  \item A qualitative expert user study that demonstrates that our dynamic problem lists offer statistically significant improvements over a strong baseline as a clinical decision support tool.
\end{itemize}
\vspace{-\topsep}

\subsection*{Generalizable Insights about Machine Learning in the Context of Healthcare}

A significant body of past work develops predictive models that can not be used in clinically useful settings due to their reliance on billing codes assigned after a patient leaves the hospital \citep{harlan_claims, He2014, Ghassemi2014, arash, sun1, Barbieri2020}. While there may be value in the technical innovations made by such work, research that acknowledges and addresses the constraints of the domain is essential to develop methods that can actually be implemented in practice. We demonstrate that recent methods for automated ICD code assignment are sufficiently performant to extract billing information in real-time for downstream modeling tasks. Although we focus on extracting problem lists for clinical decision support in this work, this finding has broader ramifications for the field. It both enables the real-time implementation of previously impracticable work and paves the way for future work to develop clinically feasible models that utilize dynamically extracted diagnosis and procedure information from clinical text.

\section{Related Work}

There has been a large body of prior work utilizing natural language processing (NLP) techniques to extract information from clinical narratives. \citet{sontag} demonstrated that unstructured clinical notes could be used to effectively identify patients with heart failure in real time. Their methods that involved data from clinical notes outperformed those using only structured data, demonstrating
the importance of effectively utilizing the rich source of information contained within the clinical narrative.

Prior work has found success predicting ICD code assignment using clinical notes within MIMIC-III and has found that deep learning techniques outperform traditional methods \citep{elhadad, caml, soa, multi_icd}. \citet{caml} augmented a convolutional model with a per-label attention mechanism and found that it led to both improved performance and greater interpretability as measured by a qualitative, expert evaluation. \citet{soa} later improved upon their model by utilizing multiple convolutions of different widths and then max-pooling across the channels before the attention mechanism. 

There has also been work done demonstrating that machine learning models can effectively leverage the unstructured clinical narrative for the prediction of clinical outcomes \citep{Ghassemi2014, nyu, analysis_attn_clinical}. \citet{analysis_attn_clinical} augmented long short-term memory networks (LSTMs) with an attention mechanism and applied it to predict clinical outcomes such as mortality and ICU readmission. However, when defining readmission, they treated both ICU readmissions and deaths as positive examples. The clinical work by \citet{harlan} has demonstrated that these are orthogonal outcomes, and thus modeling them jointly as a single outcome does not make sense from a clinical perspective. By treating them as separate outcomes in this work, we are able to independently explore the risk factors for these two distinct outcomes.

\citet{analysis_attn_clinical} also raised some questions about the interpretability of attention in their work with clinical notes, repeating the experiments introduced by \citet{attn_not} to evaluate the explanatory capabilities of attention. However, \citet{attn_not_not} explored some of the problems with their underlying assumptions and experimental setup and demonstrated that their experiment failed to fully explore their premise, and thus failed to support their claim. 

\begin{figure}[h]
  \centering
  \includegraphics[width=\linewidth]{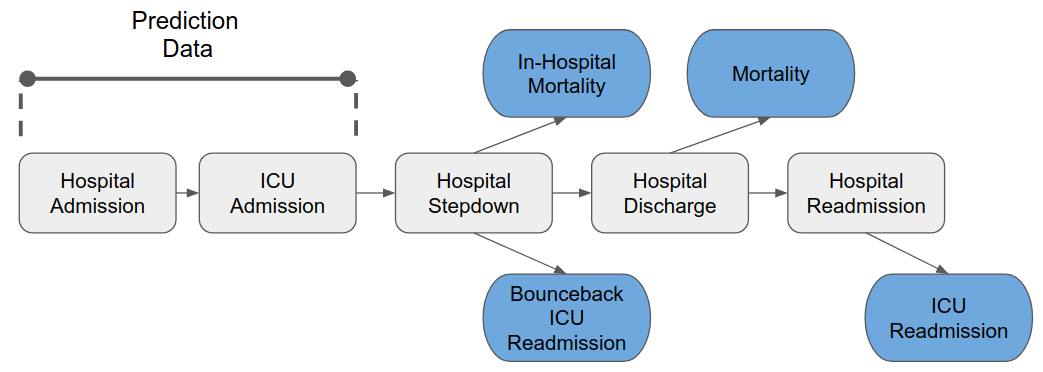}
  \caption{Outcomes explored in this work}
  \label{outcomes-figure}
\end{figure}

\section{Data and cohort}
This work is conducted using the free text notes stored in the publicly available MIMIC-III database \citep{mimic}. The database  contains de-identified clinical data for over forty thousand patients who stayed in the critical care units of the Beth Israel Deaconess Medical Center. This information was collected as part of routine clinical care and, as such, is representative of the information that would be available to clinicians in real-time. This makes the dataset well-suited for developing clinical models. 

To develop our cohort, we first filter out minors because children have different root causes for adverse medical outcomes than the general populace. We also remove patients who died while in the ICU and filter out ICU stays that are missing information regarding the time of admission or discharge. We then extract all ICU stays where the patient had at least three notes on record before the time of ICU discharge to develop a cohort with a meaningful textual history. This leaves us with $33,311$ unique patients and $45,260$ ICU stays.

For ICU readmission we extract labels for two types of readmissions, bounceback and 30 day readmisssion. Bounceback readmissions occur when a patient is discharged from the ICU and then readmitted to the ICU before being discharged from the hospital. For 30 day readmissions, we simply look at any readmission to the ICU within the 30 days following ICU discharge. For mortality, we also look at two different outcomes, in-hospital mortality and 30-day mortality. Because we use all data available at the time of ICU discharge, in-hospital mortality is constrained to mortality that occurs after ICU discharge but prior to hospital discharge. All the outcomes that we explored in this work are laid out in Figure \ref{outcomes-figure}. This provides us with a cohort with $3,413$ $(7.5\%)$ bounceback readmissions, $5,674$ $(12.5\%)$ 30-day readmissions, $3,761$ $(8.3\%)$ deaths within 30 days, and $1,898$ $(4.2\%)$ in-hospital deaths. For our experiments, we then split our cohort into training, validation, and testing splits following an 80/10/10 split and use 5-fold cross validation. We divide our cohort based on the patient rather than the ICU stay to avoid data leakage when one patient has multiple ICU stays.

We extract all clinical notes associated with a patient's hospital stay up until the time of their discharge from the ICU. The text is then preprocessed by lowercasing the text, normalizing punctuation, and replacing numerical characters and de-identified information with generic tokens. All of the notes for each patient are then concatenated and treated as a continuous sequence of text which is used as the input to all of our models. We truncate or pad all clinical narratives to 8000 tokens. This captures the entire clinical narrative for over $75\%$ of patients and we found that extending the maximum sequence length beyond that point did not lead to any further improvements in performance.

\section{Methods}

\begin{figure}[h]
  \centering
  \includegraphics[width=\linewidth]{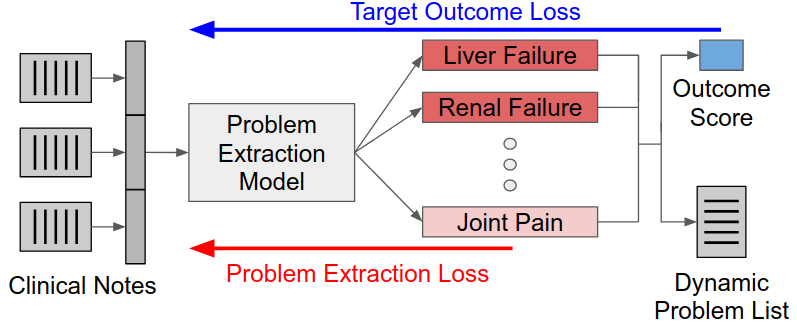}
  \caption{Overview of our proposed framework}
  \label{overview}
\end{figure}

In this work, we develop an end-to-end framework to jointly extract problems from the clinical narrative and then use those problems to predict a target outcome of interest. An overview of our framework can be seen in Figure \ref{overview}. We embed the clinical notes using learned word embeddings and then apply a convolutional attention model with a guided multi-headed attention mechanism to extract problems from the narrative. We then utilize the intermediate problem predictions to predict the target outcome. This differs from standard deep learning models because the features used for our final prediction are clearly mapped to clinically meaningful problems rather than opaque learned features. We also describe the training procedure that we develop to ensure that our problem extraction model maintains a high level of performance, something that is essential for the intermediate features to maintain their clinical significance.

\subsection{Embedding techniques}
We utilize all notes in the MIMIC-III database associated with subjects who are not in our testing set to train embeddings using the Word2Vec method \citep{w2v}. This allows for training on a greater selection of notes than if training had been limited to the training set. This training is done using the continuous bag-of-words implementation and it generates embeddings for all words that appear in at least 5 notes in our corpus. We replace out-of-vocabulary words with a randomly initialized UNK token to represent unknown words. Both 100 and 300 dimensional word embeddings were explored and early testing showed that 100 dimensional word embeddings led to better performance. 

\subsection{Target Problems}
We experiment with multiple different representations for the intermediate problems in this work. The first representation we explore are the ICD9 codes assigned to all hospital stays in our dataset. These codes are used for billing purposes and represent diagnostic and procedure information for each patient. Although prior work has found that these codes are predictive of adverse outcomes \citep{Ghassemi2014, arash, Barbieri2020}, these codes are assigned after a patient has been discharged from the hospital and, as such, directly using these codes as features in a predictive model limits the clinical utility of such a model. By instead learning to dynamically assign these codes within our framework, we can use these codes to predict the outcomes we explore using only the information available at the time of prediction.

However, the large ICD9 label space will likely hinder our frameworks's ability to effectively extract and utilize the codes. To address this, we leverage the heirarchical nature of the ICD9 taxonomy. Full ICD9 codes are represented by character strings up to $6$ characters in length where each subsequent character represents a finer grained distinction. We experiment with rolled up ICD9 codes which consist of only the first three characters of each ICD9 code to address the problem of the large label space. The rolled up codes still represent clinically meaningful procedures and conditions while substantially reducing the number of labels.

We also explore using phecodes which were developed to conduct phenome-wide association studies (PheWAS) in EHRs \citep{phewas}. Prior work demonstrated that phecodes better represent clinically meaningful phenotypes than ICD9 codes \citep{Wei_2017}. Because of this, phecodes may lead to a more clinically meaningful and predictive intermediate representations than ICD9 diagnosis codes. A mapping from ICD9 codes to phecodes already exists and can be used to extract phecodes from our dataset. Similar to ICD9 codes, we explore both full and rolled up phecodes. For every problem representation in this work, we only use codes that occur at least $50$ times in our training set to reduce label sparsity. After this filtering, there are an average of $1047.4$ full ICD diagnosis codes, $331.8$ full ICD procedure codes, $695.6$ full phecodes codes, $419.6$ rolled ICD diagnosis codes, $203.4$ rolled ICD procedure codes, and $356.0$ rolled phecodes across our $5$ folds.

\subsection{Problem extraction model}

\begin{figure}[h]
  \centering
  \includegraphics[width=.7\linewidth]{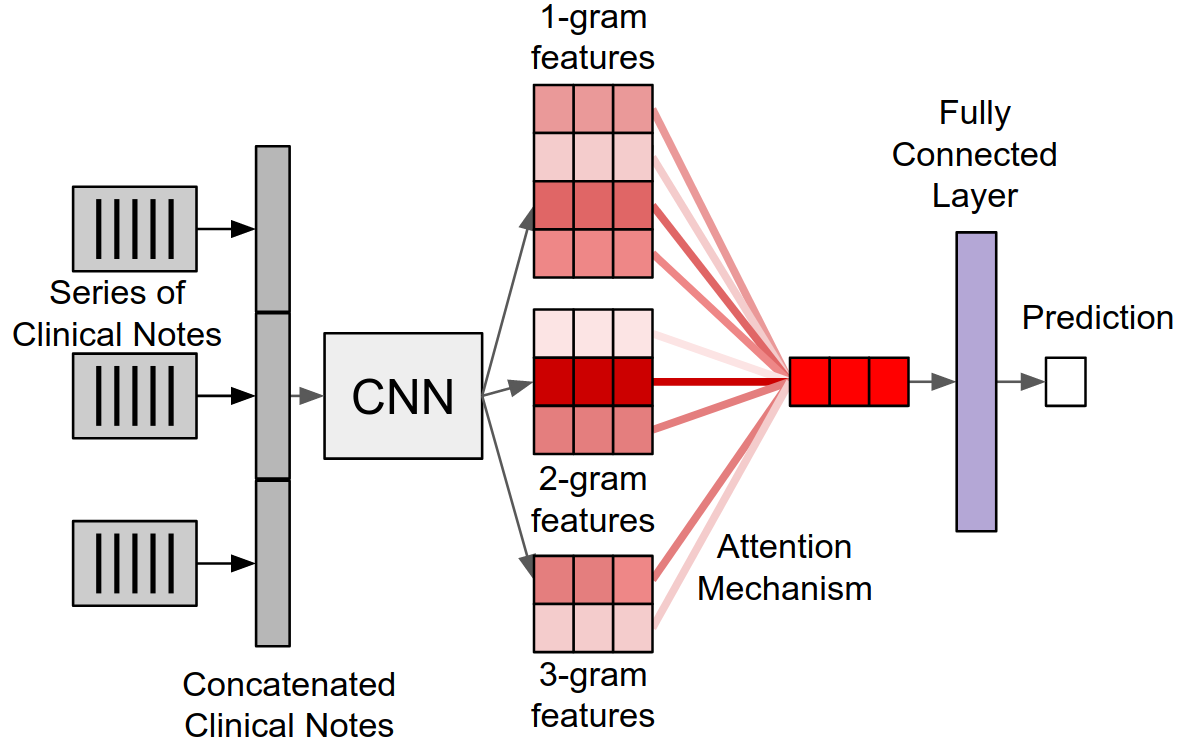}
  \caption{Illustration of our problem extraction model with a single attention mechanism shown.}
\end{figure}

The convolutional attention architecture used in this work is similar to that developed by \citet{caml} and \citet{soa} for automatic ICD code assignment. The model can be described as follows. We represent the clinical narrative as a sequence of $d_e$-dimensional dense word embeddings. Those word embeddings are then concatenated to create the matrix $\mathbf{X} = [\mathbf{x}_1;\mathbf{x}_2;...;\mathbf{x}_N]$ where $N$ is the length of the clinical narrative and ${x}_n\in \mathbb{R}^{d_e}$ is the word embedding for the $n^{th}$ word in the narrative. We then apply a convolutional neural network (CNN) to the matrix $\mathbf{X}$.

In this work, we use three convolutional filters of width 1, 2, and 3 with output dimensionality $d_f$. These filters convolve over the textual input with a stride of 1, applying the learned filters to every 1-gram, 2-gram. and 3-gram in the input. In this work, we augment the CNN with a multi-headed attention mechanism where each head is associated with a problem \citep{attn_all_you_need}. Unlike the work of \citet{caml} and \citet{soa}, we apply our attention mechanisms over multiple convolutional filters of different lengths. This allows our model to consider variable spans of text while still maintaining the straightforward interpretability of the model introduced by \citet{caml}. 

To apply the attention mechanisms, we learn a query vector, $\mathbf{q_{\ell}} \in \mathbb{R}^{d_f}$, for each problem $\ell$ that will be used to calculate the importance of the feature maps across all filters for that problem. We calculate the importance using the dot product of each feature map with the query vector. We let $\mathbf{H}\in\mathbb{R}^{d_f \times (3N)}$ be the concatenated output of our CNN and can then calculate the attention distribution over all of the feature maps simultaneously using the matrix vector product of our final feature map and the query vector as $\boldsymbol{\alpha}_{\ell}=softmax(\frac{\mathbf{H^{T}}\mathbf{q}_{\ell}}{\sqrt{d_f}})$ where $d_f$ is used as a scaling factor and $\boldsymbol{\alpha_{\ell} \in \mathbb{R}^{3N}}$ contains the score for every position across all the filters. The softmax operation is used so that the score distribution is normalized. We calculate the final representation used for classification for problem $\ell$ by taking a weighted average of all of the outputs based on their calculated weights given by $\mathbf{v}_{\ell} = \sum_{i=1}^{3N}\boldsymbol{\alpha_{\ell,i}}\mathbf{h_i}$ where $\mathbf{h}_{i}$ is the $i^{\text{th}}$ feature vector in $\mathbf{H}$ and $\mathbf{v}_{\ell}$ is the final representation used for predicting the presence of problem $\ell$.

Given the representation $\mathbf{v}_{\ell}$, we calculate the final prediction as $\hat{y_{\ell}}=\sigma(\mathbf{w^{T}}_{\ell}\mathbf{v}_{\ell} + b_{\ell})$ where $\mathbf{w}_{\ell}$ is a vector of learned weights, $\mathbf{b}_{\ell}$ is the bias term, and $\sigma$ is the sigmoid function. We train our problem extraction model by minimizing the binary cross-entropy loss function given by $\mathcal{L}_{p} = -\sum_{\ell=1}^{L}y_{\ell}\textrm{log}(\hat{y}_{\ell})+(1-y_{\ell})\textrm{log}(1-\hat{y}_{\ell})$ where $y_{\ell}$ is the ground truth label and $\hat{y}_{\ell}$ is our model's prediction for problem ${\ell}$.

\subsection{Outcome classification}
In our proposed framework, the feature vector used for the outcome prediction is
$\mathbf{s} = [s_{0};s_{1};...;s_{L-1};s_{L}]$ where $\mathbf{s} \in \mathbb{R}^{L}$ and $s_{\ell}$ is the scalar score for problem $\ell$ defined by $s_{\ell} = \mathbf{w}^{T}_{\ell}\mathbf{v}_{\ell} + b_{\ell}$ . We calculate our final prediction using this vector similarly to our intermediate problem prediction as $\hat{y_{0}}=\sigma(\mathbf{w^{T}}_{o}\mathbf{s} + b_{o})$.
Using the score for each outcome as the features for the final prediction allows for the straightforward interpretation of each feature. This differs from the standard deep learning models used in prior works where the final feature vector used for the prediction is composed of learned features that are not interpretable. We utilize this improvement to explain our model's decision making process and to develop dynamic problem lists.

To optimize the classification objective for our target outcome, we also minimize the  binary cross-entropy loss function $\mathcal{L}_{o} = -(y_{o}\textrm{log}(\hat{y}_{o})+(1-y_{o})\textrm{log}(1-\hat{y}_{o}))$
where $y_o$ is the ground truth label for our target outcome and $\hat{y}_{o}$ is our model's prediction for that outcome.


\subsection{Training procedure}
For our intermediate features to be interpretable, it is important for our problem extraction model to maintain a high level of performance. This motivates the development of our training procedure. We define a threshold for the performance of our problem extraction model and train only that component of our framework if the validation performance falls below that threshold. This ensures that we are only training the final classification layer using intermediate representations that effectively represent their corresponding problem. This also prevents our target classification objective from degrading the performance of our problem extraction model as that would harm the interpretability and clinical utility of our framework. 

Thus our final loss function $\mathcal{L}$ can be defined as
$ \mathcal{L} = \begin{cases} 
      \mathcal{L}_{o} + \mathcal{L}_{p} & \text{if } val_p \geq threshold_p \\
      \mathcal{L}_{p} & \text{if } val_p < threshold_p \\
   \end{cases}
$
where $val_p$ is the validation performance and $threshold_p$ is a pre-defined performance threshold. We measure the performance of our problem extraction model by calculating the micro-averaged Area Under the Receiver Operating Curve (AU-ROC) on the validation set and use a threshold of $0.90$ for the models trained in this work. We found this training procedure to be necessary to maintain good problem extraction performance for problem configurations that involved full codes while the configurations with rolled codes were able to maintain performance during joint training. We optimize our final loss function using the Adam optimizer \citep{adam}. Our code is made publicly available\footnote{\url{https://github.com/justinlovelace/Dynamic-Problem-Lists}} and we relegate full implementation details to the appendix.

\section{Experiments and results}

\subsection{Baselines}
To evaluate the efficacy of our proposed framework at predicting our target outcomes, we develop three strong baselines based on recent work for clinical outcome prediction using clinical text \citep{khadanga2019using, nyu, analysis_attn_clinical}. The first baseline is the convolutional model developed by \citet{CNN} for text classification. This model consists of three convolutions of width 1, 2, and 3 which are applied over the clinical narrative and then max-pooled. The three pooled representations are then concatenated and used for the final prediction.

The second baseline is similar to the model used for problem extraction in our proposed framework and is a straightforward extension of the model proposed by \citet{caml}. Unlike our problem extraction model, this baseline utilizes a single attention head and directly predicts the outcome of interest. This baseline allows us to not only compare the predictive performance of our model, but to also explore the improved interpretability that our framework provides. For our third baseline, we use a bidirectional LSTM augmented with an additive attention mechanism which was used by \cite{analysis_attn_clinical} in their work predicting clinical outcomes from notes.

\subsection{Outcome Results}

For each outcome in this work, we explore using both full and rolled ICD codes and phecodes as our intermediate problems. To gain insight into the effectiveness of each subset of codes, we also explore using only the rolled ICD diagnosis codes, rolled ICD procedure codes, and rolled phecodes. For every model, we report the mean and standard deviation across the five testing folds for the area under the Receiver Operating Curve (AU-ROC) and the area under the Precision-Recall Curve (AUC-PR) to evaluate the effectiveness of our models. The results for all of the outcomes explored in this work can be found in Table \ref{outcome-table}. 

As expected, we find that trying to use the entire set of ICD codes for our intermediate problem representation is relatively ineffective, being outperformed by at least one of our baselines across all outcomes. We also observe that this problem extends to trying to utilize the full set of phecodes. However, we find that our model is very effective when using rolled ICD codes or phecodes. When using rolled codes, we find that our proposed framework outperforms all baselines with multiple different problem configurations across all outcomes and performance metrics. 

Somewhat surprisingly, we find that using the individual subsets of codes does not lead to any loss in performance and appears to marginally improve performance. It is possible that the additional information provided by combining diagnostic and procedure codes is offset by difficulties that come from increasing the label space. We find that our framework leads to not only improved clinical utility (which we demonstrate later in this work), but also improved predictive performance.

\subsection{Problem Extraction Results}

\begin{table}
  \caption{Outcome Prediction Results }
  \label{outcome-table}
  \centering
  \begin{adjustbox}{center}
  \begin{tabular}{cccccc}
    \toprule
    Model     & Problem Set  & \multicolumn{2}{c}{In-Hospital Mortality} & \multicolumn{2}{c}{30-Day Mortality}      \\
    \midrule
    & & AU-ROC & AU-PR & AU-ROC & AU-PR\\ 
    
    \midrule[.15em]
    CNN-Max & -  & $0.852 \pm 0.015$ & $0.323 \pm 0.048$ & $0.842 \pm 0.008$ & $0.430 \pm 0.009$\\
    Conv-Attn & - &  $0.865 \pm 0.015$ & $0.330 \pm 0.038$ & $0.852 \pm 0.007$ & $0.415 \pm 0.012$\\
    LSTM-Attn & - &  $0.853 \pm 0.015$ & $0.308 \pm 0.046$ & $0.855 \pm 0.008$ & $0.431 \pm 0.007$\\
    DynPL & $\text{F-ICD}_{\text{Diag}}$ \&  $\text{F-ICD}_{\text{Proc}}$ & $0.823 \pm 0.023$ & $0.218 \pm 0.036$ & $0.821 \pm 0.012$ & $0.352 \pm 0.031$\\
    DynPL & F-Phe \&  $\text{R-ICD}_{\text{Proc}}$ & $0.837 \pm 0.047$ & $0.252 \pm 0.090$ & $0.836 \pm 0.013$ & $0.393 \pm 0.028$\\
    DynPL & $\text{R-ICD}_{\text{Diag}}$ \&  $\text{R-ICD}_{\text{Proc}}$ & $\mathbf{0.866 \pm 0.011}$ & $0.322 \pm 0.046$ & $\mathbf{0.857 \pm 0.005}$ & $\mathbf{0.438 \pm 0.012}$\\
    DynPL & R-Phe \&  $\text{R-ICD}_{\text{Proc}}$ & $\mathbf{0.865 \pm 0.016}$ & $\mathbf{0.330 \pm 0.040}$ & $\mathbf{0.855 \pm 0.006}$ & $\mathbf{0.435 \pm 0.007}$\\
    DynPL & $\text{R-ICD}_{\text{Diag}}$ & $\mathbf{\underline{0.869 \pm 0.010}}$ & $\mathbf{\underline{0.332 \pm 0.037}}$ & $0.852 \pm 0.008$ & $0.424 \pm 0.021$\\
    DynPL & $\text{R-ICD}_{\text{Proc}}$ & $0.863 \pm 0.011$ & $0.329 \pm 0.030$ & $\mathbf{0.855 \pm 0.005}$ & $\mathbf{\underline{0.443 \pm 0.011}}$\\
    DynPL & R-Phe & $\mathbf{0.867 \pm 0.014}$ & $0.327 \pm 0.040$ & $\mathbf{\underline{0.858 \pm 0.007}}$ & $\mathbf{0.440 \pm 0.021}$\\
    
    \bottomrule
    \smallskip \\
    \toprule
     Model     & Problem Set & \multicolumn{2}{c}{Bounceback Readmission} & \multicolumn{2}{c}{30-Day Readmission}      \\
    \midrule
          &  & AU-ROC & AU-PR & AU-ROC & AU-PR \\ 
    
    \midrule[.15em]
    CNN-Max & -  & $0.661 \pm 0.018$ & $0.148 \pm 0.016$ & $0.650 \pm 0.011$ & $0.212 \pm 0.018$\\
    Conv-Attn & - & $0.707 \pm 0.009$ & $0.173 \pm 0.018$ & $0.684 \pm 0.004$ & $0.235 \pm 0.017$\\
    LSTM-Attn & - &  $0.695 \pm 0.010$ & $0.154 \pm 0.009$ & $0.681 \pm 0.008$ & $0.231 \pm 0.021$\\
    DynPL & $\text{F-ICD}_{\text{Diag}}$ \& $\text{F-ICD}_{\text{Proc}}$ & $0.667 \pm 0.015$ & $0.138 \pm 0.018$ & $0.659 \pm 0.011$ & $0.213 \pm 0.016$\\
    DynPL &  F-Phe \&  $\text{R-ICD}_{\text{Proc}}$ & $0.692 \pm 0.014$ & $0.154 \pm 0.013$ & $0.669 \pm 0.006$ & $0.219 \pm 0.008$\\
    DynPL & $\text{R-ICD}_{\text{Diag}}$ \&  $\text{R-ICD}_{\text{Proc}}$ & $0.703 \pm 0.013$ & $0.168 \pm 0.021$ & $0.683 \pm 0.003$ & $0.234 \pm 0.016$\\
    DynPL & R-Phe \&  $\text{R-ICD}_{\text{Proc}}$ & $0.705 \pm 0.017$ & $0.168 \pm 0.019$ & $\mathbf{0.687 \pm 0.005}$ & $0.234 \pm 0.011$\\
    DynPL & $\text{R-ICD}_{\text{Diag}}$ & $\mathbf{\underline{0.710 \pm 0.014}}$ & $0.170 \pm 0.011$ & $\mathbf{0.688 \pm 0.004}$ & $\mathbf{0.238 \pm 0.012}$\\
    DynPL & $\text{R-ICD}_{\text{Proc}}$ & $\mathbf{0.708 \pm 0.012}$ & $\mathbf{\underline{0.178 \pm 0.019}}$ & $\mathbf{\underline{0.690 \pm 0.006}}$ & $\mathbf{\underline{0.239 \pm 0.017}}$\\
    DynPL & R-Phe & $\mathbf{\underline{0.710 \pm 0.013}}$ & $\mathbf{0.173 \pm 0.019}$ & $\mathbf{0.689 \pm 0.003}$ & $\mathbf{0.238 \pm 0.011}$\\
    \bottomrule
  \end{tabular} 
  \end{adjustbox}
  \\
  \medskip
F=Full Codes, R=Rolled Codes.  Bolded values indicate equivalent or superior performance compared to all baselines and the best performance is underlined.
\smallskip \\
  \caption{Problem Extraction Results}
  \label{icd-table}
  \centering
  \begin{adjustbox}{center}
  \begin{tabular}{x{2.2cm}x{1.56cm}x{1.56cm}x{1.56cm}x{1.56cm}x{1.56cm}x{1.56cm}x{1.56cm}x{1.56cm}}
    \toprule
    \multicolumn{1}{c}{Target Outcome} & \multicolumn{2}{c}{$\text{F-ICD}_{\text{Diag}}$ \& $\text{F-ICD}_{\text{Proc}}$} & \multicolumn{2}{c}{F-Phe \&  $\text{R-ICD}_{\text{Proc}}$}  & \multicolumn{2}{c}{$\text{R-ICD}_{\text{Diag}}$ \& $\text{R-ICD}_{\text{Proc}}$} & \multicolumn{2}{c}{R-Phe \&  $\text{R-ICD}_{\text{Proc}}$ }      \\
    \midrule
       & Micro AU-ROC & Macro AU-ROC & Micro AU-ROC & Macro AU-ROC & Micro AU-ROC & Macro AU-ROC & Micro AU-ROC & Macro AU-ROC\\ 
    
    \midrule[.15em]
    Problem Extraction  & $\mathbf{0.946 \pm}$ $\mathbf{ 0.001}$ & $\mathbf{0.887 \pm}$ $\mathbf{ 0.002}$ & $\mathbf{0.945\pm}$ $\mathbf{ 0.001}$ & $\mathbf{0.877 \pm}$ $\mathbf{ 0.002}$ & $\mathbf{0.952 \pm}$ $\mathbf{ 0.000}$ & $\mathbf{0.888 \pm}$ $\mathbf{ 0.002}$ & $\mathbf{0.952 \pm}$ $\mathbf{ 0.001}$ & $\mathbf{0.879 \pm}$ $\mathbf{ 0.003}$  \\
    Bounceback Readmission& $0.853 \pm 0.005$ & $0.753 \pm 0.005$ & $0.889 \pm 0.003$ & $0.760 \pm 0.007$ & $0.905 \pm 0.002$ & $0.754 \pm 0.009$ & $0.908 \pm 0.002$ & $0.744 \pm 0.010$  \\
    30-Day Readmission  & $0.865 \pm 0.022$ & $0.756 \pm 0.013$ & $0.891 \pm 0.009$ & $0.764 \pm 0.006$ & $0.905 \pm 0.001$ & $0.748 \pm 0.008$ & $0.908 \pm 0.002$ & $0.739 \pm 0.010$  \\
    In-Hospital Mortality  & $0.862 \pm 0.022$ & $0.738 \pm 0.014$ & $0.887 \pm 0.012$ & $0.753 \pm 0.021$ & $0.906 \pm 0.004$ & $0.754 \pm 0.007$ & $0.906 \pm 0.003$ & $0.740 \pm 0.009$  \\
    30-Day Mortality  & $0.847 \pm 0.026$ & $0.733 \pm 0.021$ & $0.893 \pm 0.011$ & $0.757 \pm 0.006$ & $0.902 \pm 0.002$ & $0.749 \pm 0.006$ & $0.902 \pm 0.002$ & $0.733 \pm 0.007$  \\
    \bottomrule
  \end{tabular}
  \end{adjustbox}
\end{table}

For our model to be interpretable, it is important for the problem extraction model to be effective. To explore the performance of our problem extraction model and the effect that the additional learning objective has on that performance, we conduct an additional experiment where we train our problem extraction model independently and compare it with the performance of our intermediate problem extraction model in our framework across all outcomes. We report results for this experiment in Table \ref{icd-table}. 

We observe that our problem extraction method is performant across all of the target outcomes in this work. However, we find that our problem extraction model is consistently more effective when using rolled codes as opposed to full sets of codes. This is understandable as the larger label space and finer grained distinctions between the codes leads to a more challenging classification problem. This reduced problem extraction performance when using the full set of codes is likely a contributing factor to the poorer target outcome performance observed when using full sets of codes.

We do observe that the addition of the target outcome objective does degrade performance when compared to a model trained exclusively on problem extraction. This degradation demonstrates the importance of our training procedure to ensure that the  intermediate problem extraction remains effective.

\subsection{Effect of End-to-End Training}
We conduct an ablation experiment to evaluate the effect of end-to-end training on our framework's performance by first training our framework only on problem extraction, freezing the problem extraction component, and then fine-tuning the final classification layer to predict the outcome of interest. We report results for this experiment in Table \ref{end-to-end} and observe a consistent decrease in performance when training the two components separately. This decrease is particularly notable for both mortality outcomes. This is likely because the feature space defined by the problems fail to represent all pertinent information from the notes and training the network end-to-end allows for some adaptation to the final outcome. For example, the frozen problem extraction model would not be incentivized to recognize the severity of problems while such information would be useful when predicting the target outcomes. 

\begin{table}
  \caption{Effect of End-to-End Training}
  \label{end-to-end}
  \centering
  \begin{adjustbox}{center}
  \begin{tabular}{cccccc}
    \toprule
    Model     & Problem Set  & \multicolumn{2}{c}{In-Hospital Mortality} & \multicolumn{2}{c}{30-Day Mortality}      \\
    \midrule
    & & AU-ROC & AU-PR & AU-ROC & AU-PR\\ 
    
    \midrule[.15em]
    DynPL & $\text{R-ICD}_{\text{Diag}}$ \& $\text{R-ICD}_{\text{Proc}}$ & $\mathbf{0.866 \pm 0.011}$ & $\mathbf{0.322 \pm 0.046}$ & $\mathbf{0.857 \pm 0.005}$ & $\mathbf{0.438 \pm 0.012}$\\
    Frozen DynPL & $\text{R-ICD}_{\text{Diag}}$ \& $\text{R-ICD}_{\text{Proc}}$ & $0.852 \pm 0.008$ & $0.254 \pm 0.032$ & $0.847 \pm 0.006$ & $0.365 \pm 0.024$\\
    DynPL & R-Phe \& $\text{R-ICD}_{\text{Proc}}$ & $\mathbf{0.865 \pm 0.016}$ & $\mathbf{0.330 \pm 0.040}$ & $\mathbf{0.855 \pm 0.006}$ & $\mathbf{0.435 \pm 0.007}$\\
    Frozen DynPL & R-Phe \& $\text{R-ICD}_{\text{Proc}}$ & $0.837 \pm 0.017$ & $0.215 \pm 0.035$ & $0.834 \pm 0.011$ & $0.322 \pm 0.032$\\
    \bottomrule
    \smallskip \\
    \toprule
     Model     & Problem Set & \multicolumn{2}{c}{Bounceback Readmission} & \multicolumn{2}{c}{30-Day Readmission}      \\
    \midrule
          &  & AU-ROC & AU-PR & AU-ROC & AU-PR \\ 
    
    \midrule[.15em]
    DynPL & $\text{R-ICD}_{\text{Diag}}$ \& $\text{R-ICD}_{\text{Proc}}$ & $\mathbf{0.703 \pm 0.013}$ & $\mathbf{0.168 \pm 0.021}$ & $\mathbf{0.683 \pm 0.003}$ & $\mathbf{0.234 \pm 0.016}$\\
    Frozen DynPL & $\text{R-ICD}_{\text{Diag}}$ \& $\text{R-ICD}_{\text{Proc}}$ & $0.698 \pm 0.011$ & $0.161 \pm 0.012$ & $0.677 \pm 0.004$ & $0.224 \pm 0.010$\\
    DynPL &  R-Phe \& $\text{R-ICD}_{\text{Proc}}$ & $\mathbf{0.705 \pm 0.017}$ & $\mathbf{0.168 \pm 0.019}$ & $\mathbf{0.687 \pm 0.005}$ & $\mathbf{0.234 \pm 0.011}$\\
    Frozen DynPL &  R-Phe \& $\text{R-ICD}_{\text{Proc}}$ & $0.700 \pm 0.008$ & $0.163 \pm 0.008$ & $0.680 \pm 0.008$ & $0.229 \pm 0.017$\\
    \bottomrule

  \end{tabular} 
  \end{adjustbox}
\end{table}

\subsection{Comparison Against Oracle}
We conduct an additional experiment to explore the effectiveness of our problem extraction model. In this experiment we train a logistic regression oracle to predict the outcomes directly from the ground truth labels derived from ICD codes. It is important to note that because ICD codes are associated with entire hospital stays in our dataset, this experiment involves using future information compared to the clinically useful application setting of our other models. Not only are ICD codes themselves unavailable at the time of ICU discharge, but the codes could represent medical problems or procedures that arise or occur later in a patient's hospital stay after the patient is discharged from the ICU.

Nevertheless, this experiment can provide some insight into the effectiveness of our problem extraction model and whether it is currently a performance bottleneck. We report results for this logistic regression oracle across two of our problem configurations in Table \ref{oracle}. We find that using the ground truth labels leads to notably improved performance compared to our framework for the readmission outcomes, but actually leads to worse performance for most of the mortality outcomes. While the improvement for readmission outcomes can likely be attributed in part to the use of future information, the improvement likely also results from the improved accuracy of the problem labels, suggesting that the efficacy of our problem extraction model is a limiting factor in our framework's performance. However, our framework is not reliant on any particular architecture for problem extraction and this experiment demonstrates that as advances continue to be made on the task of automated ICD coding, our framework will become increasingly viable. The worse performance for mortality outcomes again suggests that the problem space doesn't perfectly represent all of the relevant information contained within the notes and highlights the importance of our end-to-end training regime which allows for some adaptation to the outcome of interest.

\begin{table}
  \caption{Comparison Against Oracle}
  \label{oracle}
  \centering
  \begin{adjustbox}{center}
  \begin{tabular}{cccccc}
    \toprule
    Model     & Problem Set  & \multicolumn{2}{c}{In-Hospital Mortality} & \multicolumn{2}{c}{30-Day Mortality}      \\
    \midrule
    & & AU-ROC & AU-PR & AU-ROC & AU-PR\\ 
    
    \midrule[.15em]
    DynPL & $\text{R-ICD}_{Diag}$ \& $\text{R-ICD}_{\text{Proc}}$ & $0.866 \pm 0.011$ & $0.322 \pm 0.046$ & $\mathbf{0.857 \pm 0.005}$ & $\mathbf{0.438 \pm 0.012}$\\
    LR Oracle & $\text{R-ICD}_{Diag}$ \& $\text{R-ICD}_{\text{Proc}}$ & $\mathbf{0.875 \pm 0.015}$ & $\mathbf{0.331 \pm 0.062}$ & $0.839 \pm 0.003$ & $0.404 \pm 0.012$\\
    DynPL &  R-Phe \& $\text{R-ICD}_{\text{Proc}}$ & $\mathbf{0.865 \pm 0.016}$ & $\mathbf{0.330 \pm 0.040}$ & $\mathbf{0.855 \pm 0.006}$ & $\mathbf{0.435 \pm 0.007}$\\
    LR Oracle &  R-Phe \& $\text{R-ICD}_{\text{Proc}}$ & $0.850 \pm 0.015$ & $0.268 \pm 0.049$ & $0.818 \pm 0.010$ & $0.320 \pm 0.039$\\
    \bottomrule
    \smallskip \\
    \toprule
     Model     & Problem Set & \multicolumn{2}{c}{Bounceback Readmission} & \multicolumn{2}{c}{30-Day Readmission}      \\
    \midrule
          &  & AU-ROC & AU-PR & AU-ROC & AU-PR \\ 
    
    \midrule[.15em]
    DynPL & $\text{R-ICD}_{\text{Diag}}$ \& $\text{R-ICD}_{\text{Proc}}$ & $0.703 \pm 0.013$ & $0.168 \pm 0.021$ & $0.683 \pm 0.003$ & $0.234 \pm 0.016$\\
    LR Oracle & $\text{R-ICD}_{\text{Diag}}$ \& $\text{R-ICD}_{\text{Proc}}$ & $\mathbf{0.807 \pm 0.013}$ & $\mathbf{0.294 \pm 0.039}$ & $\mathbf{0.732 \pm 0.007}$ & $\mathbf{0.314 \pm 0.016}$\\
    DynPL &  R-Phe \& $\text{R-ICD}_{\text{Proc}}$ & $0.705 \pm 0.017$ & $0.168 \pm 0.019$ & $0.687 \pm 0.005$ & $0.234 \pm 0.011$\\
    LR Oracle &  R-Phe \& $\text{R-ICD}_{\text{Proc}}$ & $\mathbf{0.808 \pm 0.013}$ & $\mathbf{0.286 \pm 0.034}$ & $\mathbf{0.733 \pm 0.007}$ & $\mathbf{0.312 \pm 0.013}$\\
    \bottomrule

  \end{tabular} 
  \end{adjustbox}
\end{table}

\subsection{Label Integrity}
Although our framework's problem extraction performance provides a straightforward way to validate the effectiveness of our problem extraction model, it is not a perfect method due to the nature of our ground truth labels. A number of past works have demonstrated that ICD codes are an imperfect representation of ground truth phenotypes in actual clinical practice \citep{Chang2016, Benesch660, 10.2307/3768402, doi:10.1161/01.STR.30.1.56, doi:10.2105/AJPH.82.2.243, 10.1093/aje/kwh314, doi:10.1161/STROKEAHA.114.006316, doi:10.1161/CIRCOUTCOMES.113.000743, doi:10.1161/STROKEAHA.113.003408}. A common trend observed in work exploring the accuracy of ICD codes is that they have strong specificity but poorer sensitivity. In other words, a patient assigned a given code very likely has the corresponding condition, but there are likely more patients with that condition than only the patients who were assigned that ICD code. Given that our dataset contains information gathered during routine clinical care, the ICD codes we use as ground truth labels in this work likely suffer from the same problem.

Because of this complication, perfect problem extraction performance, as evaluated by using ICD codes as ground truth labels, is actually suboptimal. In such a case, the model would have learned to perfectly replicate the biases and mistakes in the ICD coding process instead of correctly identifying all of the clinical problems. We hypothesize that if our problem extraction model is effective, then there are likely some 'incorrect' predictions that count against our model in the evaluation above that are actually correct. To evaluate this hypothesis, we conduct an expert evaluation over a limited set of predictions.

Because ICD codes tend to have problems with sensitivity, most of the errors with our ICD labels should be false negatives. To evaluate whether our problem extraction model is correctly recognizing some of the problems missed by the ICD codes, we extract the $50$ most confident false positives for one of the models trained in this work and manually evaluate whether the patient actually has the corresponding problem. It is important to note that when conducting the evaluation, we are not necessarily following ICD coding standards. We are instead identifying whether the patient has the corresponding problem to explore challenges with using ICD codes to represent phenotype labels as is being done in this work and has been done in prior work \citep{rodriguez2018phenotype}. We report the results for this experiment in Table \ref{human-eval}.

\begin{table}[h]
  \caption{Expert Evaluation of 50 False Positives}
  \label{human-eval}
  \centering 
  \begin{tabular}{ccccccc}
  \toprule
     & Count & Percentage\\
     \cmidrule(r){1-3} 
    Correct Prediction & $37$ & $74\%$ \\
    Correct Label & $13$ & $26\%$ \\
    \bottomrule
  \end{tabular}
\end{table}

We observe that our hypothesis was correct and that a large majority ($74\%$) of the false positives that we extracted from our model were actually correct predictions penalized due to label inaccuracies. This demonstrates that our model is already reasonably robust to these label inaccuracies and is successfully extracting problems despite noisy labels. We also observe that the actual false positives are often well grounded in the text. For example, radiologists prioritize sensitivity over specificity when reporting observations, and we found multiple false positives resulting from radiological findings that required clinical correlation. Although there is a large body of work in ICD code classification in MIMIC \citep{caml, soa, elhadad,multi_icd}, we are the first to conduct such an analysis demonstrating the ability of our model to overcome label inconsistencies.

\section{Interpretability}
While we demonstrated that our framework is performant, its primary strength is the simplicity of interpretation that it provides. \citet{explainable} surveyed clinicians to identify aspects of explainable modeling that improve clinician's trust in machine learning models. Clinicians identified being able to understand feature importance as a critical aspect of explainability so that they can easily compare the model's decision with their clinical judgement. Clinicians expected to see both global feature importance and patient-specific importance so we explore both of those in this work. 



\subsection{Global Trends}

A large body of prior work has explored the interpretability of attention, but that exploration is typically limited to individual predictions \citep{caml, show_attend_and_tell, hu}. While that is useful, it is also important to gain an understanding of population level trends. 

By designing our frameworks such that the value for the final prediction is a linear combination of the extracted problem scores, we can simply extract the weights from the final layer of our model to gain an understanding of which problems are important. We calculated the mean and standard deviation for each problem over the five folds and present the strongest risk factors across all outcomes in Table \ref{risks}. We observe that there are a number of common risk factors between outcomes. We find that the top four risk factors for both readmission tasks were fluid disorders; puncture of vessel; renal failure; and congestive heart failure, not hypertensive. We find that urinary tract infections and pneumonia were both strong factors for mortality as well as the shared readmission risk factors of puncture of vessel and fluid disorders.

We also explored whether there were factors associated with healthy outcomes but found that even the most negative weights had a small magnitude that was insignificant given their variance. Thus our model appears to recognize a limited number of positive risk factors while the majority of the intermediate problems have little effect on the outcome. This makes it well-suited for producing clutter-free problem lists for clinicians which we explore in the next section.

\begin{table}[h]
  \caption{Risk Factors for Target Outcomes }
  \label{risks}
  \centering 
  \begin{adjustbox}{center}
  \begin{tabular}{x{7cm}c|x{7cm}c}
  \toprule
    \multicolumn{2}{c}{30-Day Mortality} & \multicolumn{2}{c}{In-Hospital Mortality}       \\
    \cmidrule(r){1-4} 
     Problem  & Weight &  Problem  & Weight\\ 
    \midrule
    Disorders of fluid, electrolyte, and acid-base balance & $0.151 \pm 0.057$ & Disorders of fluid, electrolyte, and acid-base balance & $0.189 \pm 0.027$ \\
    Puncture of vessel & $0.091 \pm 0.035$ & Urinary tract infection & $0.081 \pm 0.018$ \\
    Pneumonia & $0.078 \pm 0.026$ & Puncture of vessel & $0.079 \pm 0.060$ \\
    Urinary tract infection & $0.072 \pm 0.040$ &  Renal failure & $0.073 \pm 0.020$ \\
    Congestive heart failure; nonhypertensive & $0.067 \pm 0.016$ &  Pneumonia & $0.071 \pm 0.018$ \\

  \toprule
    \multicolumn{2}{c}{30-Day Readmission} & \multicolumn{2}{c}{Bounceback Readmission}       \\
    \cmidrule(r){1-4} 
     Problem  & Weight &  Problem  & Weight\\ 
    \midrule
     Disorders of fluid, electrolyte, and acid-base balance & $0.110 \pm 0.019$ & Disorders of fluid, electrolyte, and acid-base balance & $0.081 \pm 0.025$ \\
    Renal failure & $0.081 \pm 0.022$ & Puncture of vessel & $0.076 \pm 0.028$ \\
    Puncture of vessel & $0.072 \pm 0.019$ & Congestive heart failure; nonhypertensive & $0.059 \pm 0.014$ \\
    Congestive heart failure; nonhypertensive & $0.069 \pm 0.021$ &  Renal failure & $0.037 \pm 0.014$\\
    Other anemias & $0.069 \pm 0.061$ &  Hypertension & $0.037 \pm 0.034$\\
    
    \bottomrule
  \end{tabular}
  \end{adjustbox}
\end{table}

\begin{table}
  \caption{Dynamic Problem Lists}
  \label{dpl-high-risk}
  \centering 
  \begin{adjustbox}{center}
  \begin{tabular}{x{7.5cm}x{2cm}x{2cm}x{8.5cm}}
  \multicolumn{4}{c}{High-Risk Bounceback Readmission Patient} \\
  \toprule
    Problem &  Extraction Probability & Problem Weight & Top Two Spans of Attended Text \\
    \cmidrule(r){1-4} 
    
     Other operations of abdominal region (includes paracentesis) &  $0.950$ & $0.16$ & {[to attempt \textbf{paracentesis} again today]} \newline {[suitable for \textbf{paracentesis} was marked]} \\
    
     Chronic liver disease and cirrhosis &  $0.939$ & $0.28$ & {[to attempt \textbf{paracentesis} again today]} \newline {[suitable for \textbf{paracentesis} was marked]} \\
     
     Injection or infusion of therapeutic or prophylactic substance &  $0.838$ & $0.31$ & {[started on \textbf{tpn} plan was]} \newline {[remains on \textbf{tpn} at present]} \\
    
     Puncture of vessel &  $0.797$ & $1.00$ & {[, beir \textbf{hugger applied d/t} low  temp.;]} \newline {[reddend alovesta \textbf{cream applied id} : tmax]} \\
     
     Disorders of fluid, electrolyte, and acid-base balance &  $0.732$ & $0.92$ & {[will be \textbf{performed lft's elevated} being followed]} \newline {[pt is \textbf{jaundiced , excoriated} perianal area]} \\
     
     Septicemia &  $0.556$ & $0.15$ & {[support , \textbf{sepsis work-up} p-will]} \newline {[levofloxacin and \textbf{flagyl} po skin]} \\
     
     Ascites (non malignant) &  $0.539$ & $0.12$ & {[to attempt \textbf{paracentesis} again today]} \newline {[suitable for \textbf{paracentesis} was marked]} \\
     
     Transfusion of blood and blood components &  $0.484$ & $0.06$ & {[pt had \textbf{egd} this pm]} \newline {[rec'd \# \textbf{units} ffp with]} \\
     
     Prophylactic vaccination and inoculation against certain viral diseases &  $0.460$ & $0.06$ & {[support , \textbf{sepsis} work-up p-will]} \newline {[history of \textbf{hepatorenal} failure and]} \\
     
     Chronic ulcer of skin &  $0.450$ & $0.32$ & {[, beir \textbf{hugger applied} d/t low  temp.;]} \newline {[reddend alovesta \textbf{cream applied id} : tmax]}  \\
     
     Renal failure &  $0.404$ & $0.39$ & {[s/p now \textbf{with renal failure} reason for]} \newline {[s/p now \textbf{with renal failure} reason for]}  \\
     
     Peritonitis and retroperitoneal infections &  $0.360$ & $0.04$ & {[to attempt \textbf{paracentesis} again today]} \newline {[suitable for \textbf{paracentesis} was marked]}  \\
     
     Other anemias &  $0.359$ & $0.06$ & {[rec'd n \textbf{units} ffp with]} \newline {[rec'd n \textbf{unit} ffp with]}  \\
     
     Viral hepatitis &  $0.349$ & $0.13$ & {[status , \textbf{lactulose} prn as]} \newline {[remains on \textbf{lactulose} prn to]}  \\

     \toprule
     \multicolumn{4}{c}{Low-Risk Bounceback Readmission Patient (Truncated)} \\
  \toprule
    \cmidrule(r){1-4} 
     Diagnostic procedures on small intestine &  $0.547$ & $-0.07$ & {[presently another \textbf{endoscopy} is scheduled]} \newline {[had an \textbf{endoscopy} which revealed]}  \\
     Other anemias &  $0.284$ & $0.06$ & {[nnd unit \textbf{prbc} infusing presently]} \newline {[n unit \textbf{prbc} with initial]}  \\
     
     Diseases of esophagus &  $0.252$ & $-0.05$ & {[presently another \textbf{endoscopy} is scheduled]} \newline {[had an \textbf{endoscopy} which revealed]}  \\
     
     Effects radiation NOS &  $0.223$ & $0.06$ & {[nnd unit \textbf{prbc} infusing presently]} \newline {[n unit \textbf{prbc} with initial]}  \\

    \bottomrule
  \end{tabular}
  \end{adjustbox}
\end{table}

\begin{table}
  \caption{Baseline Attention Interpretation}
  \label{attn-high-risk}
  \centering 
  \begin{adjustbox}{center}
  \begin{tabular}{x{9.5cm}x{9.5cm}}
  \toprule
    \multicolumn{2}{c}{Highly Attended Text} \\
    \toprule
    High-Risk Bounceback Readmission Patient & Low-Risk Bounceback Readmission Patient \\
    \midrule
     {[radiology to \textbf{attempt paracentesis} again today]} & {[small amts \textbf{ice chips awaiting} nnd endoscopy]}  \\
     {[iv bid \textbf{old tap} site from]} & {[understanding of \textbf{discharge instructions} and new]}  \\
     {[planning to \textbf{do tap} this evening]} & {[daughters care \textbf{discharge instructions} reviewed with]}  \\
     {[further oozing \textbf{needs c-diff spec} pmicu nursing]} & {[fbleeding noted \textbf{discharge instructions} , pt]}  \\
     {[was d/cd \textbf{a paracentesis} was attempted]} & {[ice chips \textbf{per team neuro} : a\&oxn]}  \\
     {[of ice \textbf{chips tpn} infusing as]} & {[taking medication \textbf{discharge planning} complete with]}  \\
     {[overnight mushroom \textbf{cath draining loose} brown-green stool]} & {[scheduled for \textbf{this am- ?} nam pt]}  \\
     {[was started \textbf{on tpn} plan was]} & {[of chron's \textbf{disease} and lower]}  \\
     {[pt remains \textbf{on tpn} at present]}  & {[, denies \textbf{sob} rr nn-nn]} \\
     {[status , \textbf{lactulose} prn as]}  & {[, dry \textbf{, intact without} reddness or]} \\
     {[remains on \textbf{lactulose} prn to]}  & {[up the \textbf{clots} pt transferred]} \\
     {[re-oriented rec'ing \textbf{lactulose} po has]} & {[chron's disease \textbf{and lower gib} , now]}  \\
     {[pt given \textbf{lactulose} x n]} &  {[in the \textbf{<loc> area plan} : repeat]} \\
     {[on po \textbf{lactulose} perl ,]}  & {[given iv \textbf{erythromycin} and iv]} \\
    \bottomrule
  \end{tabular}
  \end{adjustbox}
\end{table}

\subsection{Individual Predictions }
We construct dynamic problem lists by extracting the 14 strongest problem predictions. We chose to extract 14 problems because the patients in the training fold had an average of $13.8$ codes assigned to their hospital stay so 14 problems should provide an adequate summary of the patient's state. We report these problems sorted by their extraction probability and also report the importance of each problem for the final outcome so that clinicians can easily identify what factors are driving the prediction. For the problem importance, we scale the problem weights to the range $[-1,1]$ by dividing by the problem weight with the greatest magnitude to allow for easier interpretation, and we also provide the spans of text attended to by the model to make each problem prediction. To provide a comparison using our baseline convolutional attention model, we extract the $14$ spans of text with the greatest attention weights associated with them.

We provide an example of a dynamic problem list for a patient predicted to be at high risk of bounceback readmission in Table \ref{dpl-high-risk}. From looking at the dynamic problem list, we can quickly identify the most important problems driving the risk prediction (puncture of vessel, fluid disorder, renal failure, skin ulcer, intravenous feeding, and liver disease) while understanding that the other problems are insignificant. Reporting the quantitative importance of each problem saves the clinician from having to manually filter through the long list of problems. Furthermore, the extraction probability provides a measure of uncertainty which, along with the attended text, allows clinicians to intelligently verify the model's performance. For example, renal failure is an important risk factor but has a relatively low extraction probability of $0.404$. Upon inspecting the highlighted text, the clinician can clearly observe that the extraction was accurate and the patient is suffering from that condition. It is also worth noting that in this example the problem extraction model was able to successfully recognize that the patient had bed ulcers and a platelet transfusion, both of which are not represented by the ICD labels in the dataset.

By comparison, we provide the baseline visualization from the convolutional attention model for the same patient in Table \ref{attn-high-risk}. Here, we can only observe much broader trends and there is a large degree of redundancy (e.g. paracentesis and tap refer to the same procedure). We can observe that the patient has severe liver problems from the need for paracentesis and the use of the medication, lactulose. We can also observe that the patient required intravaneous feeding from the references to total parenteral nutrition (TPN). However, there is a significant amount of redundancy and it is not clear how to meaningfully aggregate these observations to actually gain an understanding of what clinical outcomes the model is extracting and how important they are for the final outcome. Furthermore, the overview of the patient is much less comprehensive than that provided by the dynamic problem list, with all of the information extracted by the baseline being concisely aggregated into three codes (Chronic liver disease and cirrhosis, Other operations of abdominal region, and Injection or infusion of therapeutic or prophylactic substance) in the dynamic problem list that quantitatively reports the importance of those conditions.

We compare a dynamic problem list to our baseline for a low-risk bounceback patient in the same tables and find that the benefits are even more pronounced. When examining the baseline visualization we observe that the model is primarily focusing on references to discharge instructions which don't actually convey any clinically meaningful information. Similarly, the other phrases attended to do not seem to convey any important medical information. On the other hand, the dynamic problem list for the low-risk patient still effectively extracts clinical conditions (that the patient had an esophageal disease, was anemic, etc.) and then concludes that the extracted conditions do not warrant concern. This clearly demonstrates to a clinician that the model is still effectively extracting the patient's clinical condition, but that it judges that condition to be safe. This transparency is important for the clinician to be able to trust that the model is effective.

\section{Qualitative Expert User Study}
While we have argued for the improved utility of our framework compared against recent work within the domain, it is important to verify that claim by conducting a user study with medical experts. For example, it may be possible that while our framework is sound in theory, the problem extraction stage is sufficiently noisy to render the extracted problem lists useless in practice. To examine the utility of our framework, we recruited four medical experts and conducted a user study where our experts evaluated the utility of our dynamic problem list and the baseline interpretation method. Three of our experts are currently practicing physicians while one is an MD-PhD student with one year of medical school remaining. Two of the medical experts are co-authors who were involved in some parts of the development of this work while the other two had no involvement with our work beyond taking part in the user study.

We conducted our user study by randomly sampling $25$ ICU stays from the test set of one of our $30$ day readmission models. Because of the imbalanced nature of our dataset, we sample $10$ stays from the top $5\%$ of predicted risks and sample the other $15$ stays from the remaining ICU stays. This ensures that we evaluate our framework for both high risk patients and patients that are representative of the general patient population. We then provided each of our expert reviewers with the clinical notes associated with each patient and instructed them to briefly review them to gain an understanding of the patient's medical condition. We then presented them with our dynamic problem list and the baseline attention extraction along with the predicted readmission risk and the reviewers evaluated both methods independently using the Likert Scale seen in Figure \ref{likert}. 

We report the results for this study in Table \ref{user-study} and compute the statistical significance for two comparisons. We examine the relationship between the two interpretation methods using a two-tailed paired t-test and also explore whether the dynamic problem list is meaningfully better than a neutral rating using a two-tailed one sample t-test. The first comparison allows us to examine whether our method is an improvement over the baseline while the second allows us to evaluate whether the medical expert's judged our method favorably. We observe that every expert found our framework to be more effective than the baseline method and the difference was statistically significant for all but one expert. Additionally, every expert found the problem list to be meaningfully better than a neutral rating by a statistically significant margin. By contrast, two of our experts rated the baseline worse than neutral and none of the experts rated it to be better than neutral by a statistically significant margin. When averaging the scores for each patient across all experts, we find that our method received a rating of $3.66$ on average compared to $2.85$ for the baseline method, a meaningful improvement over both the baseline ($p<0.001$) and a neutral rating ($p<0.001$). These improvements are still significant even when limiting the evaluation to the two external experts to account for potential biases from the experts who were familiar with this work. While a much more stringent evaluation would need to be conducted (such as a randomized controlled trial) before implementing our method in practice, this preliminary qualitative evaluation is promising and more rigorous evaluations are left to future work.

\begin{table}[h]
  \caption{Likert Scale}
  \label{likert}
  \centering 
   \begin{adjustbox}{center}
  \begin{tabular}{x{3cm}x{3cm}x{3cm}x{3cm}x{3cm}}
  \toprule
     \multicolumn{5}{C{15cm}}{The list effectively identifies and presents relevant medical factors for evaluating readmission risk for this patient.				}				\\
     \midrule
     Strongly Disagree & Disagree & Neutral & Agree &  Strongly Agree \\
     1 & 2 & 3 & 4 & 5\\
     
    \bottomrule
  \end{tabular}
  \end{adjustbox}
\end{table}

\begin{table}[h]
  \caption{User Study}
  \label{user-study}
  \centering 
  \begin{adjustbox}{center}
  \begin{tabular}{ccccccc}
  \toprule
         & \multicolumn{4}{c}{Medical Expert}  \\
    & 1 & 2 & 3 & 4 & Average & Average of External Experts \\ 
    \midrule
     Convolutional Attention & $3.13$ & $2.52$ & $2.52$ & $3.32$ & $2.85$ & $2.92$\\
     Dynamic Problem List & $\mathbf{4.08}$ & $\mathbf{3.48}$ & $\mathbf{3.52}$ & $\mathbf{3.56}$ & $\mathbf{3.66}$ & $\mathbf{3.52}$ \\
     DynPL $>$ Conv-Attn & $p<0.005$ & $p<0.005$ & $p<0.01$ & $p=0.110$ & $p<0.001$ & $p<0.005$\\
     DynPL $>$ Neutral & $p<0.001$ & $p<0.05$ & $p<0.05$ & $p< 0.01$  & $p< 0.001$ & $p< 0.01$\\
    \bottomrule
  \end{tabular}
  \end{adjustbox}
\end{table}

\section{Limitations and Future Work}
We did not make the problem extraction architecture a large focus of this work and instead used a model representative of the recent state-of-the-art. In the future, we intend to improve upon the problem extraction module in our framework. In particular, we intend to explore whether we can utilize pre-trained language models to improve our problem extraction and downstream performance given their recent success across a wide variety of tasks both outside of and within the clinical domain \citep{devlin-etal-2019-bert, alsentzer-etal-2019-publicly}. In this work, we augmented our problem extraction module with a linear layer for its simplicity of interpretation and found that it led to strong performance. However, incorporating our problem extraction module into a more sophisticated model could potentially lead to meaningful improvements in performance and we intend to pursue this in future work. We would also like to extend this framework to other outcomes of clinical interest such as sepsis or the onset of intubation to evaluate its ability to generalize beyond the outcomes examined in this work.

\section{Conclusion}
In this work we develop a framework to extract outcome-specific problem lists from the clinical narrative while jointly predicting the likelihood of that outcome. We demonstrate that our framework is both more performant and more transparent than competitive baselines. Although there is a large body of work that has utilized billing information for clinical modeling, we are the first to demonstrate that it can be dynamically extracted in clinically useful settings to develop performant models. We also conduct a novel analysis to demonstrate that our problem extraction model is robust to labeling errors found in real-world clinical data. By reducing the final decision to a linear model that uses interpretable intermediate problems, we easily extract risk factors associated with the outcomes studied in this work. We also utilize this improved transparency to produce dynamic problem lists which were viewed more favorably than a competitive baseline according to an expert user study.


\bibliography{references}

\appendix
\section*{Appendix A.}
The output dimensionality of all of our convolutional filters is set to $64$. We apply dropout with a probability of $0.2$ after the embedding layer and apply it with a probability of $0.3$ after the convolutional layer and before every linear layer. For our LSTM model we use 128 hidden units and similarly apply dropout with a probability of $0.2$ after the embedding layer and apply it with a probability of $0.3$ before the final prediction. All of our models were trained with an effective batch size of 32 (gradient accumulation was necessary for some of the larger models) using a learning rate of $0.001$ with the Adam optimizer and are trained using early stopping based on their performance on the validation set. We train each model for a maximum of 100 epochs and stop training early when the AU-ROC for our target outcome has not improved for 10 epochs with stable problem extraction performance. We then evaluate the model with the best validation performance as measured by the AU-ROC on the test set. All of our hyperparameters were tuned based on validation performance. 

\end{document}